# Optimal Time Bounds for Approximate Clustering

Ramgopal R. Mettu　　C. Greg Plaxton
Department of Computer Science
University of Texas at Austin
Austin, TX 78712, U.S.A.
{ramgopal, plaxton}@cs.utexas.edu

## Abstract

Clustering is a fundamental problem in unsupervised learning, and has been studied widely both as a problem of learning mixture models and as an optimization problem. In this paper, we study clustering with respect the $k$-median objective function, a natural formulation of clustering in which we attempt to minimize the average distance to cluster centers. One of the main contributions of this paper is a simple but powerful sampling technique that we call *successive sampling* that could be of independent interest. We show that our sampling procedure can rapidly identify a small set of points (of size just $O(k \log n/k)$) that summarize the input points for the purpose of clustering. Using successive sampling, we develop an algorithm for the $k$-median problem that runs in $O(nk)$ time for a wide range of values of $k$ and is guaranteed, with high probability, to return a solution with cost at most a constant factor times optimal. We also establish a lower bound of $\Omega(nk)$ on any randomized constant-factor approximation algorithm for the $k$-median problem that succeeds with even a negligible (say $\frac{1}{100}$) probability. The best previous upper bound for the problem was $\tilde{O}(nk)$, where the $\tilde{O}$-notation hides polylogarithmic factors in $n$ and $k$. The best previous lower bound of $\Omega(nk)$ applied only to deterministic $k$-median algorithms. While we focus our presentation on the $k$-median objective, all our upper bounds are valid for the $k$-means objective as well. In this context our algorithm compares favorably to the widely used $k$-means heuristic, which requires $O(nk)$ time for just one iteration and provides no useful approximation guarantees.

## 1 Introduction

Clustering is a fundamental problem in unsupervised learning that has found application in many problem domains. Approaches to clustering based on learning mixture models as well as minimizing a given objective function have both been well-studied [1, 2, 3, 4, 5, 9]. In recent years, there has been significant interest in developing clustering algorithms that can be applied to the massive data sets that arise in problem domains such as bioinformatics and information retrieval on the World Wide Web. Such data sets pose an interesting challenge in that clustering algorithms must be robust as well as fast. In this paper, we study the *k-median problem* and obtain an algorithm that is time optimal for most values of $k$ and with high probability produces a solution whose cost is within a constant factor of optimal.

A natural technique to cope with a large set of unlabeled data is to take a random sample of the input in the hopes of capturing the essence of the input and subsituting the sample for the original input. Ideally we hope that the sample size required to capture the relevant information in the input is significantly less than the original input size. However, in many situations naive sampling does not always yield the desired reduction in data. For example, for the problem of learning Gaussians, this limitation manifests itself in the common assumption that the mixing weights are large enough so that a random sample of the data will capture a nonnegligible amount of the mass in a given Gaussian. Without this assumption, the approximation guarantees of recent algorithms for learning Gaussians [1, 4] no longer hold.

A major contribution of our work is a simple yet powerful sampling technique that we call *successive sampling*. We show that our sampling technique is an effective data reduction technique for the purpose of clustering in the sense it captures the essence of the input with a very small subset (just $O(k \log(n/k))$, where $k$ is the number of clusters) of the points. In fact, it is this property of our sampling technique that allows us to develop an algorithm for the $k$-median problem that has a running time of $O(nk)$ for $k$ between $\log n$ and $n/\log^2 n$ and, with high probability, produces a solution with cost within a constant factor of optimal.

Given a set of points and associated interpoint distances, let the *median* of the set be the point in the set that minimizes the weighted sum of distances to all other points in the set. (Remark: The median is essentially the discrete analog of the centroid, and is also called the *medoid* [10].) We study a well-known clustering problem where the goal is to partition $n$ weighted points into $k$ sets such that the sum, over all points $x$, of the weight of $x$ multiplied by the distance from $x$ to the median of set containing $x$ is minimized. This clustering problem is a variant of the classic $k$-median problem; the $k$-median problem asks us to mark $k$ of the points such that the sum over all points $x$ of the weight of $x$ times the distance from $x$ to the nearest marked point is minimized. It is straightforward to see that the optimal objective function values for the $k$-median problem and its clustering variant are equal, and furthermore that we can convert a solution to the $k$-median problem into an equal-cost solution to its clustering variant in $O(nk)$ time. We establish a lower bound of $\Omega(nk)$ time on any randomized constant-factor approximation algorithm for either the $k$-median problem or its clustering variant. Therefore, any constant-factor approximation algorithm for the $k$-median problem implies a constant-factor approximation algorithm with the same asymptotic time complexity for the clustering variant. For this reason, we focus only on the $k$-median problem in developing our upper bounds.

It is interesting to note that algorithms for the $k$-median problem can be used for a certain model-based clustering problem as well. The recent work of Arora and Kannan [1] formulates an approximation version of the problem of learning arbitrary Gaussians. Given points from a Gaussian mixture, they study the problem of identifying a set of Gaussians whose log-likelihood is within a constant factor of the log-likelihood of the original mixture. Their solution to this learning problem is to reduce it to the $k$-median problem and apply an existing constant-factor approximation algorithm for $k$-median. Thus, our techniques may also have applicability in model-based clustering.

In this paper, we restrict our attention to the *metric* version of the $k$-median problem, in which the $n$ input points are assumed to be drawn from a metric space. That is, the interpoint distances are nonnegative, symmetric, satisfy the triangle inequality, and the distance between points $x$ and $y$ is zero if and only if $x = y$. For the sake of brevity, we write "$k$-median problem" to mean "metric $k$-median problem" throughout the remainder of the paper. It is well-known that the $k$-median problem is *NP*-hard; furthermore, it is known to be *NP*-hard to achieve an approximation ratio better than $1 + \frac{2}{e}$ [8]. Thus, we focus our attention on developing a $k$-median algorithm that produces a solution with cost within a constant factor of optimal.

### 1.1 Comparison to $k$-means

Even before the hardness results mentioned above were established, heuristic approaches to clustering such as the $k$-means heuristic were well-studied (see, e.g., [5, 10]). The $k$-means heuristic is commonly used in practice due to ease of implementation, speed, and good empirical performance. Indeed, one iteration of the $k$-means heuristic requires just $O(nk)$ time [5]; typical implementations of the $k$-means heuristic make use of a small to moderate number of iterations.

However, it is easy to construct inputs with just a constant number of points that, for certain initializations of $k$-means, yield solutions whose cost is not within any constant factor of the optimal cost. For example, suppose we have 5 unit-weight points in $\mathbb{R}^2$ where three points are colored blue and two are colored red. Let the blue points have coordinates $(0, 1),(0, 0)$, and $(0, -1)$, and let the red points have coordinates $(-D, 0)$ and $(D, 0)$. For $k = 3$, the optimal solution has cost 1, whereas the $k$-means heuristic, when initialized with the blue points, converges to a solution with cost $2D$ (the blue points). Since $D$ can be arbitrarily large, in this case the $k$-means heuristic does not produce a solution within any constant factor of optimal. Indeed, a variety of heuristics for initializing $k$-means have been previously proposed, but no such initialization procedure is known to ensure convergence to a constant-factor approximate solution.

The reader may wonder whether, by not restricting the $k$ output points to be drawn from the $n$ input points, the $k$-means heuristic is able to compute a solution of substantially lower cost than would otherwise be possible. The reduction in the cost is at most a factor of two since given a $k$-means solution with cost $C$, it is straightforward to identify a set of $k$ input points with cost at most $2C$.

The $k$-means heuristic typically uses an objective function that sums squared distances rather than distances. The reader may wonder whether this variation leads to a substantially different optimization problem. It is straightforward to show that squaring the distances of a metric space yields a distance function that is "near-metric" in the sense that all of the properties of a metric space are satisfied except that the triangle inequality only holds to within a constant factor (2, in this case). It is not difficult to show that all of our upper bounds hold, up to constant factors, for such near-metric spaces. Thus, if our algorithm is used as the initialization procedure for $k$-means, the cost of the resulting solution is guaranteed to be within a constant factor of optimal. Our algorithm is particularly well-suited for this purpose because its running time, being comparable to that of a single iteration of $k$-means, does not dominate the overall running time.

## 1.2 Our Results

Before stating our results we introduce some useful terminology that we use throughout this paper. Let $U$ denote the set of all points in a given instance of the $k$-median problem; we assume that $U$ is nonempty. A *configuration* is a nonempty subset of $U$. An *m-configuration* is a configuration of size at most $m$. (Remark: An $m$-configuration is simply a set of $m$ cluster centers.) For any points $x$ and $y$ in $U$, let $w(x)$ denote the nonnegative weight of $x$, let $d(x, y)$ denote the distance between $x$ and $y$, and let $d(x, X)$ be defined as $\min_{y \in X} d(x, y)$. The *cost* of any configuration $X$, denoted $cost(X)$, is defined as $\sum_{x \in U} d(x, X) \cdot w(x)$. We denote the minimum cost of any $m$-configuration by $OPT_m$. For brevity, we say that an $m$-configuration with cost at most $a \cdot OPT_k$ is an *(m, a)-configuration*. A $k$-median algorithm is *(m, a)-approximate* if it produces an $(m, a)$-configuration. A $k$-median algorithm is *a-approximate* if it is $(k, a)$-approximate. In light of the practical importance of clustering in the application areas mentioned previously, we also consider the given interpoint distances and point weights in our analysis. Let $R_d$ denote the ratio of the diameter of $U$ (i.e., the maximum distance between any pair of points in $U$) to the minimum distance between any pair of distinct points in $U$. Let $R_w$ denote the ratio of the maximum weight of any point in $U$ to the minimum nonzero weight of any point in $U$. (Remark: We can assume without loss of generality that at least one point in $U$ has nonzero weight since the problem is trivial otherwise.) Let $r_d = 1 + \lfloor \log R_d \rfloor$ and $r_w = 1 + \lfloor \log R_w \rfloor$.

Under the standard assumption that the point weights and interpoint distances are polynomially bounded, our main result is a randomized $O(1)$-approximate $k$-median algorithm that runs in $O(n(k + \log n) + k^2 \log^2 n)$ time. Then, we only need $k = \Omega(\log n)$ and $k = O(\frac{n}{\log^2 n})$ to obtain a time bound of $O(nk)$. Our algorithm succeeds *with high probability*, that is, for any positive constant $\xi$, we can adjust constant factors in the definition of the algorithm to achieve a failure probability less than $n^{-\xi}$.

We also establish a matching $\Omega(nk)$ lower bound on the running time of any randomized $O(1)$-approximate $k$-median algorithm with a nonnegligible success probability (e.g., at least $\frac{1}{100}$), subject to the requirement that $R_d$ exceeds $n/k$ by a sufficiently large constant factor relative to the desired approximation ratio. To obtain tight bounds for the clustering variant, we also prove an $\Omega(nk)$ time lower bound for any $O(1)$-approximate algorithm, but we only require that $R_d$ be a sufficiently large constant relative to the desired approximation ratio. Additionally, our lower bounds assume only that $R_w = O(1)$. Due to space constraints, we have omitted the details of this result here; the complete proofs can be found in [11].

The key building block underlying our $k$-median algorithm is a novel sampling technique that we call "successive sampling". The basic idea is to take a random sample of the points, set aside a constant fraction of the $n$ points that are "close" to the sample, and recurse on the remaining points. We show that this technique rapidly produces a configuration whose cost is within a constant factor of optimal. Specifically, for the case of uniform weights, our successive sampling algorithm yields a $(k \log(n/k), O(1))$-configuration with high probability in $O(n \max\{k, \log n\})$ time.

In addition to this sampling result, our algorithms rely on an extraction technique due to Guha *et al.* [6] that uses a black box $O(1)$-approximate $k$-median algorithm to compute a $(k, O(1))$-configuration from any $(m, O(1))$-assignment. The black box algorithm that we use is the linear-time deterministic online median algorithm of Mettu and Plaxton [12].

In developing our randomized algorithm for the $k$-median problem we first consider the case of uniform weights, where $R_w = r_w = 1$. For this special case we provide a randomized algorithm running in $O(n \max\{k, \log n\})$ time subject to the constraint $r_d \log \frac{n}{k} = O(n)$. The uniform-weights algorithm is based directly on the two building blocks discussed above: We apply the successive sampling algorithm to obtain $(k \log(n/k), O(1))$-configuration and then use the extraction technique to obtain a $(k, O(1))$-configuration. We then use this algorithm to develop a $k$-median algorithm for the case of arbitrary weights. Our algorithm begins by partitioning the $n$ points into $r_w$ power-of-2 weight classes and applying the uniform-weights algorithm within each weight class (i.e., we ignore the differences between weights belonging to the same weight class, which are less than a factor of 2 apart). The union of the $r_w$ $k$-configurations thus obtained is an $(r_w k, O(1))$-configuration. We then make use of our extraction technique to obtain a $(k, O(1))$-configuration from this $(r_w k, O(1))$-configuration.

## 1.3 Problem Definitions

Without loss of generality, throughout this paper we consider a fixed set of $n$ points, $U$, with an associated distance function $d : U \times U \to \mathbb{R}$ and an associated nonnegative demand function $w : U \to \mathbb{R}$. We assume that $d$ is a metric, that is, $d$ is nonnegative, symmetric, satisfies the triangle inequality, and $d(x, y) = 0$ iff $x = y$. For a configuration $X$ and a set of points $Y$, we let $cost(X, Y) = \sum_{x \in Y} d(x, X) \cdot w(x)$ and we let $cost(X) = cost(X, U)$. For any set of points $X$, we let $w(X)$ denote $\sum_{x \in X} w(x)$.

We define an *assignment* as a function from $U$ to $U$. For any assignment $\tau$, we let $\tau(U)$ denote the set $\{\tau(x) \mid x \in U\}$. We refer to an assignment $\tau$ with $|\tau(U)| \leq m$ as a *m-assignment*. Given an assignment $\tau$, we define the cost of $\tau$, denoted $c(\tau)$, as $\sum_{x \in U} d(x, \tau(x)) \cdot w(x)$. It is straightforward to see that for any assignment $\tau$, $cost(\tau(U)) \leq$ $c(\tau)$. For brevity, we say that an assignment $\tau$ with $|\tau(U)| \leq m$ and cost at most $a \cdot OPT_k$ is an $(m, a)$-***assignment***. For an assignment $\tau$ and a set of points $X$, we let $c(\tau, X) = \sum_{x \in X} d(x, \tau(x)) \cdot w(x)$.

The input to the $k$-median problem is $(U, d, w)$ and an integer $k$, $0 < k \leq n$. Since our goal is to obtain a $(k, O(1))$-configuration, we can assume without loss of generality that all input points have nonzero weight. We note that for all $m$, $0 < m \leq n$, removing zero weight points from an $m$-configuration at most doubles its cost. To see this, consider an $m$-configuration $X$; we can obtain an $m$-configuration $X'$ by replacing each zero weight point with its closest nonzero weight point. Using the triangle inequality, it is straightforward to see that $cost(X') \leq 2cost(X)$. This argument can be used to show that any minimum-cost set of size $m$ contained in the set of nonzero weight input points has cost at most twice $OPT_m$. We also assume that the input weights are scaled such that the smallest weight is 1; thus the input weights lie in the range $[1, R_w]$. For output, the $k$-median problem requires us to compute a minimum-cost $k$-configuration. The ***uniform weights*** $k$-median problem is the special case in which $w(x)$ is a fixed real for all points $x$. The output is also a minimum-cost $k$-configuration.

### 1.4 Previous Work

The first $O(1)$-approximate $k$-median algorithm was given by Charikar *et al.* [3]. Subsequently, there have been several improvements to the approximation ratio (see, e.g., [2] for results and citations). In this section, we focus on the results that are most relevant to the present paper; we compare our results with other recent randomized algorithms for the $k$-median problem. The first of these results is due to Indyk, who gives a randomized $(O(k), O(1))$-approximate algorithm for the uniform weights $k$-median problem [7] that runs in $\tilde{O}(nk/\delta^2)$ time, where $\delta$ is the desired failure probability.

Thorup [15] gives randomized $O(1)$-approximate algorithms for the $k$-median, $k$-center, and facility location problems in a graph. For these problems, we are not given a metric distance function but rather a graph on the input points with $m$ positively weighted edges from which the distances must be computed; all of the algorithms in [15] run in $\tilde{O}(m)$ time. Thorup [15] also gives an $\tilde{O}(nk)$ time randomized constant-factor approximation algorithm for the $k$-median problem that we consider. As part of this $k$-median algorithm, Thorup gives a sampling technique that also consists of a series of sampling steps but produces an $(O((k \log^2 n)/\varepsilon), 2 + \varepsilon)$-configuration for any positive real $\varepsilon$ with $0 < \varepsilon < 0.4$, but is only guaranteed to succeed with probability $1/2$.

For the data stream model of computation, Guha *et al.* [6] give a single-pass $O(1)$-approximate algorithm for the $k$-median problem that runs in $\tilde{O}(nk)$ time and requires $O(n^\varepsilon)$ space for a positive constant $\varepsilon$. They also establish a lower bound of $\Omega(nk)$ for deterministic $O(1)$-approximate $k$-median algorithms.

Mishra *et al.* [13] show that in order to find a $(k, O(1))$-configuration, it is enough to take a sufficiently large sample of the input points and use it as input to a black-box $O(1)$-approximate $k$-median algorithm. To compute a $(k, O(1))$-configuration with an arbitrarily high constant probability, the required sample size is $\tilde{O}(R_d^2 k)$. In the general case, the size of the sample may be as large as $n$, but depending on the diameter of the input metric space, this technique can yield running times of $o(n^2)$ (e.g., if the diameter is $o(n^2/k)$).

## 2 Successive Sampling

Our first result is a successive sampling algorithm that constructs an assignment that has cost $O(OPT_k)$ with high probability. We make use of this algorithm to develop our uniform weights $k$-median algorithm. (Remark: We assume arbitrary weights for our proofs since the arguments generalize easily to the weighted case; furthermore, the weighted result may be of independent interest.) Informally speaking, the algorithm works in sampling steps. In each step we take a small sample of the points, set aside a constant fraction the weight whose constituent points are each close to the sample, and recurse on the remaining points. Since we eliminate a constant fraction of the weight at each sampling step, the number of samples taken is logarithmic in the total weight. We are able to show that using the samples taken, it is possible to construct an assignment whose cost is within a constant factor of optimal with high probability. For the uniform weights $k$-median problem, our sampling algorithm runs in $O(n \max\{k, \log n\})$ time. (We give a $k$-median algorithm for the case of arbitrary weights in Section 5.)

Throughout the remainder of this paper, we use the symbols $\alpha$, $\beta$, and $k'$ to denote real numbers appearing in the definition and analysis of our successive sampling algorithm. The value of $\alpha$ and $k'$ should be chosen to ensure that the failure probability of the algorithm meets the desired threshold. (See the paragraph preceding Lemma 3.3 for discussion of the choice of $\alpha$ and $k'$.) The asymptotic bounds established in this paper are valid for any choice of $\beta$ such that $0 < \beta < 1$.

We also make use of the following definitions:

- A ***ball*** $A$ is a pair $(x, r)$, where the ***center*** $x$ of $A$ belongs to $U$, and the ***radius*** $r$ of $A$ is a nonnegative real.

- Given a ball $A = (x, r)$, we let $Points(A)$ denote the set $\{y \in U \mid d(x, y) \leq r\}$. However, for the sake

of brevity, we tend to write $A$ instead of $Points(A)$. For example, we write "$x \in A$" and "$A \cup B$" instead of "$x \in Points(A)$" and "$Points(A) \cup Points(B)$", respectively.

- For any set $X$ and nonnegative real $r$, we define $Balls(X, r)$ as the set $\cup_{x \in X} A_x$ where $A_x = (x, r)$.

## 2.1 Algorithm

The following algorithm takes as input an instance of the $k$-median problem and produces an assignment $\sigma$ such that with high probability, $c(\sigma) = O(cost(X))$ for any $k$-configuration $X$.

Let $U_0 = U$, and let $S_0 = \emptyset$. While $|U_i| > \alpha k'$:

- Construct a set of points $S_i$ by sampling (with replacement) $\lfloor \alpha k' \rfloor$ times from $U_i$, where at each sampling step the probability of selecting a given point is proportional to its weight.

- For each point in $U_i$, compute the distance to the nearest point in $S_i$.

- Using linear-time selection on the distances computed in the previous step, compute the smallest real $\nu_i$ such that $w(Balls(S_i, \nu_i)) \geq \beta w(U_i)$. Let $C_i = Balls(S_i, \nu_i)$.

- For each $x$ in $C_i$, choose a point $y$ in $S_i$ such that $d(x, y) \leq \nu_i$ and let $\sigma(x) = y$.

- Let $U_{i+1} = U_i \setminus C_i$.

Note that the loop terminates since $w(U_{i+1}) < w(U_i)$ for all $i \geq 0$. Let $t$ be the total number of iterations of the loop. Let $C_t = S_t = U_t$. By the choice of $C_i$ in each iteration and the loop termination condition, $t$ is $O(\log(w(U)/k'))$. For the uniform demands $k$-median problem, $t$ is simply $O(\log(n/k'))$. From the first step it follows that $|\sigma(U)|$ is $O(tk')$.

The first step of the algorithm can be performed in $O(nk')$ time over all iterations. In each iteration the second and third steps can be performed in time $O(|U_i| k')$ by using a (weighted) linear time selection algorithm. For the uniform demands $k$-median problem, this computation requires $O(nk')$ time over all iterations. The running times of the third and fourth steps are negligible. Thus, for the uniform demands $k$-median problem, the total running time of the above algorithm is $O(nk')$.

## 3 Analysis of Successive Sampling

The goal of this section is to establish that, with high probability, the output $\sigma$ of our successive sampling algorithm has cost $O(OPT_k)$. We formalize this statement in Theorem 1 below; this result is used to analyze the algorithms of Sections 4 and 5. The proof of the theorem makes use of Lemma 3.3, established in Section 3.1, and Lemmas 3.5 and 3.9, established in Section 3.2.

**Theorem 1** *With high probability, $c(\sigma) = O(cost(X))$ for any $k$-configuration $X$.*

*Proof:* The claim of Lemma 3.3 holds with high probability if we set $k' = \max\{k, \log n\}$ and $\alpha$ and $\beta$ appropriately large. The theorem then follows from Lemmas 3.3, 3.5, and 3.9. ∎

Before proceeding, we give some intuition behind the proof of Theorem 1. The proof consists of two main parts. First, Lemma 3.3 shows that with high probability, for $i$ such that $0 \leq i \leq t$, the value $\nu_i$ computed by the algorithm in each iteration is at most twice a certain number $\mu_i$. We define $\mu_i$ to be the minimum real for which there exists a $k$-configuration $X$ contained in $U_i$ with the property that a certain constant fraction, say $\frac{3}{4}$, of the weight of $U_i$ is within distance $\mu_i$ from the points of $X$. We note that $\mu_i$ can be used in establishing a lower bound on the cost of an optimal $k$-configuration for $U_i$. By the definition of $\mu_i$, for any $k$-configuration $Y$, a constant fraction, say $\frac{1}{4}$, of the weight of $U_i$ has distance at least $\mu_i$ from the points in $Y$. To prove Lemma 3.3, we consider an associated balls-in-bins problem. For each $i$, $1 \leq i \leq t$, we consider a $k$-configuration $X$ that satisfies the definition of $\mu_i$ and for each point in $X$, view the points in $U_i$ within distance $\mu_i$ as a weighted bin. Then, we view the random samples in the first step of the sampling algorithm as ball tosses into these weighted bins. We show that with $O(k)$ such ball tosses, a high constant fraction of the total weight of the bins is covered with high probability. Since the value of $\nu_i$ is determined by the random samples, it is straightforward to conclude that $\nu_i$ is within twice $\mu_i$.

It may seem that Theorem 1 follows immediately from Lemma 3.3, since for each $i$, we can approximate $\mu_i$ within a factor of 2 with $\nu_i$, and any optimal $k$-configuration can be charged a distance of at least $\mu_i$ for a constant fraction of the weight in $U_i$. However, this argument is not valid since for $j > i$, $U_j$ is contained in $U_i$; thus an optimal $k$-configuration could be charged $\mu_i$ and $\mu_j$ for the same point. For the second part of the proof of Theorem 1 we provide a more careful accounting of the cost of an optimal $k$-configuration. Specifically, in Section 3.2, we exhibit $t$ mutually disjoint sets with which we are able to establish a valid lower bound on the cost of an optimal $k$-configuration. That is, for each $i$, $1 \leq i \leq t$, we exhibit a subset of $U_i$ that has a constant fraction of the total weight of $U_i$ and for which an optimal $k$-configuration must be charged a distance of at least $\mu_i$. Lemma 3.9 formalizes this statement and proves a lower bound on the cost of an op-

### 3.1 Balls and Bins Analysis

We have omitted the proofs of the lemmas in this section due to space considerations; the complete proofs can be found in [11]. We provide the lemma statements so that the reader can gain a sense for the proof of Lemma 3.3. We begin by bounding the failure probability of a simpler family of random experiments related to the well-known coupon collector problem. For any positive integer $m$ and any nonnegative reals $a$ and $b$, let us define $f(m, a, b)$ as the probability that more than $am$ bins remain empty after $\lceil b \rceil$ balls are thrown at random (uniformly and independently) into $m$ bins. Techniques for analyzing the coupon collector problem (see. e.g., [14]) can be used to obtain sharp estimates on $f(m, a, b)$. However, the following simple upper bound is sufficient for our purposes.

**Lemma 3.1** *For any positive real $\varepsilon$, there exists a positive real $\lambda$ such that for all positive integers $m$ and any real $b \geq m$, we have $f(m, \varepsilon, \lambda b) \leq e^{-b}$.*

We now develop a weighted generalization of the preceding lemma. For any positive integer $m$, nonnegative reals $a$ and $b$, and $m$-vector $v = (r_0, \ldots, r_{m-1})$ of nonnegative reals $r_i$, we define define $g(m, a, b, v)$ as follows. Consider a set of $m$ bins numbered from 0 to $m - 1$ where bin $i$ has associated weight $r_i$. Let $R$ denote the total weight of the bins. Assume that each of $\lceil b \rceil$ balls is thrown independently at random into one of the $m$ bins, where bin $i$ is chosen with probability $r_i/R$, $0 \leq i < m$. We define $g(m, a, b, v)$ as the probability that the total weight of the empty bins after all of the balls have been thrown is more than $aR$.

**Lemma 3.2** *For any positive real $\varepsilon$ there exists a positive real $\lambda$ such that for all positive integers $m$ and any real $b \geq m$, we have $g(m, \varepsilon, \lambda b, v) \leq e^{-b}$ for all $m$-vectors $v$ of nonnegative reals.*

For the remainder of this section, we fix a positive real $\gamma$ such that $\beta < \gamma < 1$. For $0 \leq i \leq t$, let $\mu_i$ denote a nonnegative real such that there exists a $k$-configuration for which the following properties hold: (1) the total weight of all points $x$ in $U_i$ such that $d(x, X) \leq \mu_i$ is at least $\gamma w(U_i)$; (2) the total weight of all points $x$ in $U_i$ such that $d(x, X) \geq \mu_i$ is at least $(1 - \gamma)w(U_i)$. (Note that such a $\mu_i$ is guaranteed to exist.) Lemma 3.3 below establishes the main probabilistic claim used in our analysis of the algorithm of Section 2.1. We note that the lemma holds with high probability by taking $k' = \max\{k, \lceil \log n \rceil\}$ and $\alpha$ and $\beta$ appropriately large.

**Lemma 3.3** *For any positive real $\xi$, there exists a sufficiently large choice of $\alpha$ such that $\nu_i \leq 2\mu_i$ for all $i$, $0 \leq i \leq t$, with probability of failure at most $e^{-\xi k'}$.*

### 3.2 Upper and Lower Bounds on Cost

In this section we provide an upper bound on the cost of the assignment $\sigma$ as well a lower bound on the cost of an optimal $k$-configuration. Lemmas 3.4 and 3.5 establish the upper bound on $c(\sigma)$, while the rest of the section is dedicated to establishing the lower bound on the cost of an optimal $k$-configuration. Again, we have omitted the proofs of Lemmas 3.4, 3.6, 3.7, and 3.8 due to space considerations. We provide the lemma statements so that the reader can gain a sense for the proofs of Lemmas 3.5 and 3.9.

**Lemma 3.4** *For all $i$ such that $0 \leq i \leq t$, $c(\sigma, C_i) \leq \nu_i w(C_i)$.*

**Lemma 3.5**

$$c(\sigma) \leq \sum_{0 \leq i \leq t} \nu_i w(C_i)$$

*Proof:* Since the sets $C_i$, $0 \leq i \leq t$, form a partition of $U$ and by Lemma 3.4, we have that $c(\sigma) = \sum_{0 \leq i \leq t} c(\sigma, C_i) \leq \sum_{0 \leq i \leq t} \nu_i w(C_i)$. ∎

We now focus on establishing a lower bound on the cost of an optimal $k$-configuration. Throughout the remainder of this section we fix an arbitrary $k$-configuration $X$. For all $i$ such that $0 \leq i \leq t$, we let $F_i$ denote the set $\{x \in U_i \mid d(x, X) \geq \mu_i\}$, and for any integer $m > 0$, we let $F_i^m$ denote $F_i \setminus (\cup_{j>0} F_{i+jm})$ and we let $G_{i,m}$ denote the set of all integers $j$ such that $0 \leq j \leq t$ and $j$ is congruent to $i$ modulo $m$.

**Lemma 3.6** *Let $i$ be an integer such that $0 \leq i \leq t$ and let $Y$ be a subset of $F_i$. Then $w(F_i) \geq (1 - \gamma)w(U_i)$ and $cost(X, Y) \geq \mu_i w(Y)$.*

**Lemma 3.7** *For all integers $\ell$ and $m$ such that $0 \leq \ell \leq t$ and $m > 0$,*

$$cost(X, \cup_{i \in G_{\ell,m}} F_i^m) \geq \sum_{i \in G_{\ell,m}} \mu_i w(F_i^m).$$

For the remaining lemmas in this section, we let $r$ denote $\left\lceil \log_{(1-\beta)} \frac{1-\gamma}{3} \right\rceil$.

**Lemma 3.8** *For all $i$ such that $0 \leq i \leq t$, $w(F_i^r) \geq \frac{w(F_i)}{2}$.*

**Lemma 3.9** *For any $k$-configuration $X$,*

$$cost(X) \geq \frac{1-\gamma}{2r} \sum_{0 \leq i \leq t} \mu_i w(C_i).$$

*Proof:* Let $\ell = \arg\max_{0 \leq \ell < r} \{\sum_{i \in G_{\ell,r}} w(F_i^r)\}$ and fix a $k$-configuration $X$. Then $cost(X)$ is at least

$$
\begin{aligned}
cost(X, \cup_{i \in G_{\ell,r}} F_i^r) &\geq \sum_{i \in G_{\ell,r}} \mu_i w(F_i^r) \\
&\geq \frac{1}{r} \sum_{0 \leq i \leq t} \mu_i w(F_i^r) \\
&\geq \frac{1}{2r} \sum_{0 \leq i \leq t} \mu_i w(F_i) \\
&\geq \frac{1-\gamma}{2r} \sum_{0 \leq i \leq t} \mu_i w(U_i) \\
&\geq \frac{1-\gamma}{2r} \sum_{0 \leq i \leq t} \mu_i w(C_i),
\end{aligned}
$$

where the first step follows from Lemma 3.7, the second step follows from averaging and the choice of $\ell$, the third step follows from Lemma 3.8, the fourth step follows from Lemma 3.6, and the last step follows since $C_i \subseteq U$. ∎

## 4 Uniform Weights

In this section we use the sampling algorithm of Section 2, a black-box $k$-median algorithm and a modified version of the algorithm of Guha *et al.* [6] that we call Modified-Small-Space to obtain a fast $k$-median algorithm for the case of uniform weights. We note that algorithm Modified-Small-Space and the accompanying analysis is a slight generalization of results obtained by Guha *et al.* [6]. We omit the description and proof of correctness of algorithm Modified-Small-Space; a complete discussion can be found in [11]. Informally speaking, algorithm Modified-Small-Space works in two phases. First, we use an $(a, O(1))$-approximate $k$-median algorithm on the input to compute $\ell$ $(a, O(1))$-configurations. Then, we construct a new $k$-median problem instance from these $(a, O(1))$-configurations and use an $O(1)$-approximate $k$-median algorithm to compute a $k$-configuration. We are able to show that this $k$-configuration is actually a $(k, O(1))$-configuration.

We obtain our uniform weights $k$-median algorithm by applying our sampling algorithm in Step 2 of algorithm Modified-Small-Space and the deterministic online median algorithm of Mettu and Plaxton [12] in Step 4. We set the parameter $\ell$ of algorithm Modified-Small-Space to 1 and parameter $k'$ of our sampling algorithm to $\max\{k, \log n\}$. By Theorem 1, the output of our sampling algorithm is an $(m, O(1))$-assignment with high probability, where $m = O(\max\{k, \log n\} \log(n/k))$. The online median algorithm of Mettu and Plaxton [12] is also an $O(1)$-approximate $k$-median algorithm. By the properties of algorithm Modified-Small-Space [11], it can be shown that the resulting $k$-median algorithm is $O(1)$-approximate with high probability.

We now analyze the running time of the above algorithm on inputs with uniform weights. The time required to compute the output assignment $\sigma$ in Step 2 is $O(n \max\{k, \log n\})$. We note that the weight function required in Step 3 of Modified-Small-Space can be computed during the execution of the sampling algorithm without increasing its running time. The deterministic online median algorithm of Mettu and Plaxton [12] requires $O(|\sigma(U)|^2 + |\sigma(U)| r_d)$ time. The total time taken by the algorithm is therefore

$$
\begin{aligned}
&O(nk' + |\sigma(U)|^2 + |\sigma(U)| r_d) \\
&= O(nk' + k'^2 \log^2(n/k) + r_d k' \log(n/k)) \\
&= O(nk' + r_d k' \log(n/k)),
\end{aligned}
$$

where the first step follows from the analysis of our sampling algorithm for the case of uniform weights. By the choice of $k'$, the overall running time is $O((n + r_d \log(n/k)) \max\{k, \log n\})$. Note that if $k = \Omega(\log n)$ and $r_d \log(n/k) = O(n)$, this time bound simplifies to $O(nk)$.

## 5 Arbitrary Weights

The uniform-weights $k$-median algorithm developed in Sections 2 and 4 is $O(1)$-approximate for the $k$-median problem with arbitrary weights. However, the time bound established for the case of uniform weights does not apply to the case of arbitrary weights because the running time of the successive sampling procedure is slightly higher in the latter case. (More precisely, the running time of the sampling algorithm of Section 2 is $O(nk' \log \frac{w(U)}{k'})$ for the case of arbitrary weights.) In this section, we use the uniform-weight algorithm developed in Sections 2 and 4 to develop a $k$-median algorithm for the case of arbitrary weights that is time optimal for a certain range of $k$.

We now give a precise description of our $k$-median algorithm. Let $\mathcal{A}$ be the uniform weights $k$-median algorithm of Sections 2 and 4, and let $\mathcal{B}$ be an $O(1)$-approximate $k$-median algorithm.

- Compute sets $B_i$ for $0 \leq i < r_w$ such that for all $x \in B_i$, $2^i \leq w(x) \leq 2^{i+1}$.

- For $i = 0, 1 \ldots r_w - 1$: Run $\mathcal{A}$ with $B_i$ as the set of input points, $d$ as the distance function, $2^{i+1}$ as the fixed weight, and the parameter $k' = \max\{k, \lceil \log n \rceil\}$; let $Z_i$ denote the output. Let $\phi_i$ denote the assignment induced by $Z_i$, that is, $\phi_i(x) = y$ iff $y$ is in $Z_i$ and $d(x, Z_i) = d(x, y)$. For a point $x$, if $x \in Z_i$, let $w_{\phi_i}(x) = w(\phi_i^{-1}(x))$, otherwise let $w_{\phi_i}(x) = 0$.

- Let $\phi$ be the assignment corresponding to the union of the assignments $\phi_i$ defined in the previous step, and

let $w_\phi$ denote the weight function corresponding to the union of the weight functions $w_{\phi_i}$. Run $\mathcal{B}$ with $\phi(U)$ as the set of input points, $d$ as the distance function, and $w_\phi$ as the weight function. Output the resulting $k$-configuration.

Note that in the second step, $k'$ is defined in terms of $n$ (i.e., $|U|$) and not $|B_i|$. Thus, the argument of the proof of Theorem 1 implies that $\mathcal{A}$ succeeds with high probability in terms of $n$. Assuming that $r_w$ is polynomially bounded in $n$, with high probability we have that every invocation of $\mathcal{A}$ is successful.

We now observe that the above algorithm corresponds to algorithm Modified-Small-Space with the parameter $\ell$ is set to $r_w$, the uniform weights algorithm of Section 4 is used in step 2 of Small-Space, and the online median algorithm of Mettu and Plaxton [12] is used in step 4 of Small-Space. Thus, as in the previous section, the analysis of algorithm Modified-Small-Space implies that the output of $\mathcal{B}$ is a $(k, O(1))$-configuration with high probability.

We now discuss the running time of the above algorithm. It is straightforward to compute the sets $B_i$ in $O(n)$ time. Our uniform weights $k$-median algorithm requires $O((|B_i| + r_d \log \frac{|B_i|}{k})k')$ time to compute $Z_i$, so the time required for all invocations of $\mathcal{A}$ is

$$
\begin{aligned}
&O\left(\sum_{0 \leq i < r_w} (|B_i| + r_d \log(|B_i|/k))k'\right) \\
&= O\left(r_w\left(\frac{nk'}{r_w} + r_d k' \log\left(\frac{n}{kr_w}\right)\right)\right) \\
&= O\left(\left(n + r_d r_w \log \frac{n}{kr_w}\right)k'\right).
\end{aligned}
$$

(The first step follows from the fact that the sum is maximized when $|B_i| = n/r_w$.) Note that each weight function $w_{\phi_i}$ can be computed in $O(|B_i| k)$ time; it follows that $w_\phi$ can be computed in $O(nk)$ time. We employ the online median algorithm of [12] as the black-box $k$-median algorithm $\mathcal{B}$. Since $|\phi(U)|$ is at most $kr_w$, the time required for the invocation of $\mathcal{B}$ is $O((kr_w)^2 + kr_w r_d)$. It follows that the overall running time of the algorithm is as stated.

## 6 Concluding Remarks

In this paper, we have presented a constant-factor approximation algorithm for the $k$-median problem that runs in optimal $\Theta(nk)$ time if $\log n \leq k \leq \frac{n}{\log^2 n}$. If we use our algorithm as an initialization procedure for $k$-means, our analysis guarantees that the cost of the output of $k$-means is within a constant factor of optimal. Preliminary experimental work [11] suggests that this approach to clustering yields improved practical performance in terms of running time and solution quality.